\documentclass[12pt]{article}
\usepackage{amsmath}

\begin{document}

\begin{center}
{\Large \bf Anti-self-dual Riemannian metrics without\\[2mm] Killing
vectors,
can they be realized on K3?}\\[4mm]
{\large \bf A A Malykh$^1$, Y Nutku$^2$ and
M B Sheftel$^{1,2}$}\\[3mm]
$^1$ Department of Higher Mathematics, North Western State
Technical University, Millionnaya St. 5, 191186, St. Petersburg,
Russia
\\ $^2$ Feza G\"{u}rsey Institute, PO Box 6, Cengelkoy,
81220 Istanbul, Turkey \vspace{1mm}
\\ E-mail: specarm@online.ru, nutku@gursey.gov.tr,
sheftel@gursey.gov.tr
\end{center}

\vspace{5mm}

\noindent  {\bf Abstract} \\
Explicit Riemannian metrics with Euclidean signature and anti-self
dual curvature that do not admit any Killing vectors are
presented. The metric and the Riemann curvature scalars are
homogenous functions of degree zero in a single real potential and
its derivatives. The solution for the potential is a sum of
exponential functions which suggests that for the choice of a
suitable domain of coordinates and parameters it can be the metric
on a compact manifold. Then, by the theorem of Hitchin, it could
be a class of metrics on $K3$, or on surfaces whose universal
covering is $K3$.

 \vspace{5mm}



 \vspace{1cm}

We present Riemannian metrics with anti-self-dual curvature that
admit no Killing vectors. Our motivation to study this problem has
been $K3$ which is the most important gravitational instanton
\cite{ahs}-\cite{egh}. It is necessary that the metric on $K3$
should not admit any continuous symmetries. Our solutions do
satisfy this criterion which is necessary but not sufficient since
$K3$ is a {\it compact} $4$-dimensional Riemannian manifold. In
this note all our considerations will be local and we shall not
discuss the global problem of compactness. The metric
\begin{equation}
d s^2 =  u_{i\bar{k}} \, d\zeta^i d\bar\zeta^k \label{metr}
\end{equation}
must be hyper-K\"ahler. It has been over a century since Kummer
\cite{kummer} introduced $K3$ as a quartic surface in $CP^3$ and
half a century since Calabi \cite{calabi} pointed out that the
K\"ahler potential satisfies the elliptic complex Monge-Amp\`ere
equation
\begin{equation}
u_{1\bar 1}u_{2\bar 2} - u_{1\bar 2}u_{\bar 1 2}=1 \label{ma}
\end{equation}
hereafter to be referred to as $CMA_2$. Then the metric has
vanishing first Chern class \cite{ssc} and therefore satisfies the
Euclidean Einstein field equations.

Yau \cite{yau} has given an existence and uniqueness proof but so
far there are no explicit exact solutions of $CMA_2$ that do not
admit any continuous symmetries. Another testament to the
difficulties encountered in dealing with complex Monge-Amp\`ere
equations lies in the fact that its homogeneous version replaces
Laplace's equation as the fundamental equation governing functions
of many complex variables \cite{cln}. The principal difficulty in
constructing solutions of $CMA_2$ that would describe $K3$ lies in
the requirement that the K\"ahler metric (\ref{metr}) must not
admit any Killing vectors. In the language of differential
equations such solutions are known as non-invariant solutions of
(\ref{ma}). Recently we suggested that the method of group
foliation \cite{lie,vessiot,ovs} can serve as a regular tool for
finding non-invariant solutions of non-linear partial differential
equations. Group foliation was carried out for $CMA_2$ and the
Boyer-Finley equations in \cite{ns} and \cite{msw} respectively
using their infinite symmetry groups \cite{bw}.

However, in this note we shall adopt a different approach which
turned out to be fruitful specifically for $CMA_2$, to find an
explicit metric without any Killing vectors that has
anti-self-dual curvature. We emphasize at the outset that the
class of solutions we are considering here is not the full set of
solutions of $CMA_2$. In our approach we start with the
Mason-Newman \cite{mw} Lax pair and supplement the Lax equations
with two more linear equations such that $CMA_2$ emerges as an
algebraic compatibility condition. The would be Baker-Akhiezer
function in the standard Lax approach is now regarded as a complex
potential. Choosing symmetry characteristics \cite{olv} of $CMA_2$
for the real and imaginary parts of this potential we arrive at an
over-determined set of {\it linear} equations satisfied by one
real potential. This system is the image of $CMA_2$ supplemented
by some differential constraints after performing a Legendre
transformation. Its solution gives exact solutions of the
Euclidean Einstein equations with anti-self-dual Riemann curvature
$2$-form. Here we shall present only the final results and
postpone the detailed derivation to a future publication, a
preliminary account of which can be found in \cite{ms}.

We use the Euclidean Newman-Penrose formalism \cite{g1}, \cite{an}
to write the metric in the form
\begin{equation}
d s^2 =  l \otimes \bar l + \bar l \otimes l +
  m \otimes \bar m + \bar m \otimes m  \label{np}
\end{equation}
where the co-frame $\omega^a = \{ l, \bar l, m, \bar m \}$ is
given by
\begin{eqnarray} l&=& \frac{1}{v \left[ C ( C^2 -|A|^2
)\right]^{1/2} } \left[ C ( C d z^1 + B d z^2 ) + \bar A ( C d
\bar z^1 + \bar B d \bar z^2) \right]
\\ m&=& \frac{\left( C^2 -| A |^2 \right)^{1/2}}{v \, C^{1/2}} d z^2
\end{eqnarray}
and $ A, B, C, v$ are {\it a priori} functions of all coordinates.
The first three are expressed in terms of $v$
\begin{eqnarray}
A & = & v^2 + v_1^2 - i v v_2, \nonumber \\
B & = & v_2 v_{\bar 1} - i v ( v_1 - v_{\bar 1} ) \label{abdnew} \\
C & = & v^2 + |v_1|^2  \nonumber
\end{eqnarray}
where $v$ is a real-valued potential. Then the anti-self-duality
equations and therefore the Euclidean Einstein field equations
reduce to a system of over-determined {\it linear} equations
\begin{eqnarray}
v_{1\bar 1}+v &=&0 \nonumber\\
v_{11}+v-iv_2 &=&0 \nonumber\\
v_{1\bar 2}+i(v_{\bar 1}-v_1)&=&0 \label{10}\\
 v_{2\bar 2}+i(v_{\bar 2}-v_2)&= &0 \nonumber
\end{eqnarray}
which together with their complex conjugates make up $6$ real
equations. It is interesting to note that by virtue of (\ref{10})
we have $ A_{\bar 1} = - i B, \; C_1 = i B $ in terms of the
original potentials. We further note that there are no first order
equations implied by (\ref{10}). Since the conditions for
invariant solutions of a differential equation are first order
equations, the general solution of the system (\ref{10}) must
correspond to non-invariant solutions of $CMA_2$. This implies
that there are no Killing vectors in the metric.

It can be verified directly that the system (\ref{abdnew}),
(\ref{10}) gives an anti-self-dual Riemann curvature $2$-form
\begin{equation}
\Omega^a_{\;b} = - ^* \Omega^a_{\; b}, \hspace{1cm} \Omega^a_{\;
b} = \frac{1}{2} R^a_{\;bcd} \; \omega^c \wedge \omega^d
\label{anti}
\end{equation}
where $*$ is the Hodge star operator. Ricci-flatness follows by
virtue of the first Bianchi identity.

We shall consider here a particular solution of (\ref{10}) which
can be given in a finite form, so that it does not contain
infinite series or an integral. This is
\begin{eqnarray}
\!\!\! &\!\!\! & v = \sum_{j=-\infty}^{\infty} \exp \left\{2\,{\rm
Im}\! \left(\frac{}{} \left[\alpha_j^2(s_j^2+1)+1 \right]z^2
\right) \right\} \left\{ \frac{}{} \right. \label{solut}
\\ \!\!\! &\!\!\! &\exp\!\left[\frac{}{} 2s_j {\rm Re} (\alpha_j  z^1 ) \right]\!
{\rm Re}\!\left\{\frac{}{}\! D_j \exp\!\left[ 2 i\,\Bigl[{\rm
Im}\bigl(\alpha_j z^1 \bigr) - 2s_j {\rm Re}(\alpha_j^2 z^2)
\Bigr] \right]  \right\}  \nonumber
\\ \!\!\! &\!\!\! & \left. +
\exp\!\left[\frac{}{}\!\! -2 s_j {\rm Re}(\alpha_j z^1) \right]\!
{\rm Re}\!\left\{\frac{}{}\!\! E_j \exp\!\left[2 i\,\Bigl[{\rm
Im}\bigl(\alpha_j z^1 \bigr) + 2s_j {\rm Re}(\alpha_j^2 z^2)
\Bigr] \right]\! \right\}  \,
 \right\} \nonumber
\end{eqnarray}
where $\alpha_j,D_j, E_j$ are arbitrary complex constants and
$s_j=\sqrt{1-1/|\alpha_j|^2}$. The general solution of the linear
system (\ref{10}) can be given by the corresponding infinite
series, or by an integral representation for the case of the
continuous spectrum with $\alpha_j$ changed to $\alpha$, $s_j$ to
$s=\sqrt{1-1/|\alpha|^2}$, $D_j, E_j$ to $D(\alpha,\bar\alpha),
E(\alpha,\bar\alpha)$ respectively and the sum changed to a double
integral with respect to $\alpha,\bar\alpha$.

The locus of possible singularities of the curvature scalars is
given by the first order partial differential equation
\begin{equation}
 v \left[ \frac{}{} (v_1 - v_{\bar 1} )^2  +| v_2 |^2 \frac{}{}
 \right] - i v^2 ( v_2 - v_{\bar 2} )
 - i ( v_2 v_{\bar 1}^2- v_{\bar 2} v_1^2 )
  = 0
 \label{sing}
\end{equation}
which are also singularities of the metric. Imposing equation
(\ref{sing}) in addition to the system of equations (\ref{10})
results in further relations between the metric coefficients
$$ A = \lambda C, \qquad B = \mu C $$
where $\lambda$ and $\mu$ are arbitrary complex constants. Further
analysis of the compatibility conditions arising from (\ref{10})
and (\ref{sing}) gives the general form of the singular solution
\begin{eqnarray}
& & v = {\rm Re}\!\left\{ D \exp \left\{ \frac{}{} \alpha(s+1) z^1
+\bar\alpha(s-1)\bar z^1
 \right. \right. \label{finsol}\\
& & \left. \left. - i \! \left[\alpha^2(s+1)^2+1  \right]z^2 + i\!
\left[\bar\alpha^2(s-1)^2+1 \right]\bar z^2 \! \frac{}{} \right\}
\! \right\} \nonumber
\end{eqnarray}
where
\[\lambda=-\frac{\alpha}{\bar \alpha}\,,\quad \mu=2i\alpha s,\quad
s=\sqrt{1-\frac{1}{|\alpha|^2}}  \] and $D$ is an arbitrary
complex constant. We conclude that curvature singularities {\it
related to the choice of} $v$ will arise if and only if it is
given by (\ref{finsol}).

Any solution for $v$ of the form (\ref{solut}) when substituted
into (\ref{abdnew}) gives us an explicit form of the metric
\begin{eqnarray}
ds^2 &=& \frac{1}{ v^2 (C^2 - |A|^2 ) } \left[ \frac{}{} A(C d z^1
+ B d z^2)^2 +\bar A(C d \bar z^1+  \bar B d\bar z^2)^2\right.
\label{newmetr}
\\ & & \left.\mbox{}+\frac{1}{C}\,( C^2 + | A |^2) | C d z^1 + B d z^2 |^2
\right] + \frac{1}{ v^2 C} (C^2 -|A|^2) | d z^2 |^2 \nonumber
\end{eqnarray}
which is an exact solution of the anti-self-duality equations
(\ref{anti}) and therefore the Einstein field equations with
Euclidean signature. The fact that it admits no Killing vectors
follows from the non-trivial dependence of $v$ on all four
coordinates provided we keep a minimum of two terms in the sum
(\ref{solut}).

We shall not discuss whether, or not our solution describes a
compact $4$-manifold. All our analysis has been local and given a
metric in a local coordinate chart as in (\ref{newmetr}),
compactness is always an open question. The property of
compactness depends on the range of coordinates that we may assign
to the local coordinates. We have not done that, but the fact that
the metric coefficients and curvature scalars are homogeneous
functions of degree zero in the potential $v$ and its derivatives,
together with the presence of exponentials in the potential
suggests that the metric could well be made compact by choosing a
suitable domain of coordinates and parameters.

Assuming compactness, by virtue of its anti-self-dual curvature
property, our solution saturates Hitchin's bound $|\tau|\le
(2/3)\chi$ \cite{hitchin} between the Euler characteristic $\chi$
and the Hirzebruch signature $\tau$. By Hitchin's theorem
\cite{hitchin} we know that only $K3$, and surfaces whose
universal covering is $K3$ have this property.

\end{document}